\documentclass[aps,prl,twocolumn,superscriptaddress]{revtex4-1}   

\pdfoutput=1    % ensure pdflatex processing for arXiv

\usepackage{graphicx,color}  % needed for figures
\usepackage{dcolumn}   % needed for some tables
\usepackage{bm}        % for math
\usepackage{amssymb}   % for math
\usepackage{subfigure}
\usepackage{lipsum}   
\usepackage[heightadjust=all,valign=t]{floatrow}
\usepackage{fr-subfig}
\usepackage{sidecap}
\usepackage{comment}
\usepackage{mathrsfs}
\usepackage{soul}
\usepackage{comment} 
\usepackage{hyperref}
\usepackage{xcolor}

\newcommand{\ket}[1]{|#1\rangle}

\providecommand{\ignore}[1]{}

\usepackage{hyperref}
\hypersetup{colorlinks=true, pdfstartview=FitV, linkcolor=blue, citecolor=black, plainpages=false, pdfpagelabels=true, urlcolor=blue}
\usepackage[all]{hypcap}

\newcommand{\beginsupplement}{
    \renewcommand{\thesubsection}{\Roman{subsection}}
    \renewcommand{\theequation}{S\arabic{equation}}
    \setcounter{equation}{0}
    \renewcommand{\thefigure}{S\arabic{figure}}
    \setcounter{figure}{0}
    \renewcommand{\thetable}{S\arabic{table}}
    \setcounter{table}{0}
    \renewcommand{\theHtable}{Supplement.\thetable}
    \renewcommand{\theHfigure}{Supplement.\thefigure}
}

\begin{document}

\title{Quasiparticle Poisoning of Superconducting Qubits from Resonant Absorption of Pair-breaking Photons}

\author{C. H. Liu} 
\affiliation{Department of Physics, University of Wisconsin-Madison, Madison, Wisconsin 53706, USA}

\author{D. C. Harrison} 
\affiliation{Intelligence Community Postdoctoral Research Fellowship Program, Department of Physics, University of Wisconsin-Madison, Madison, Wisconsin 53706, USA}

\author{S. Patel} 
\affiliation{Department of Physics, University of Wisconsin-Madison, Madison, Wisconsin 53706, USA}

\author{C. D. Wilen}
\affiliation{Department of Physics, University of Wisconsin-Madison, Madison, Wisconsin 53706, USA}

\author{O. Rafferty} 
\affiliation{Department of Physics, University of Wisconsin-Madison, Madison, Wisconsin 53706, USA}

\author{A. Shearrow}
\affiliation{Department of Physics, University of Wisconsin-Madison, Madison, Wisconsin 53706, USA}

\author{A. Ballard}
\affiliation{Department of Physics, Syracuse University, Syracuse, New York 13244, USA}

\author{V. Iaia}
\affiliation{Department of Physics, Syracuse University, Syracuse, New York 13244, USA}

\author{J. Ku}
\affiliation{Department of Physics, Syracuse University, Syracuse, New York 13244, USA}

\author{B. L. T. Plourde}
\affiliation{Department of Physics, Syracuse University, Syracuse, New York 13244, USA}

\author{R. McDermott}
\email[]{rfmcdermott@wisc.edu}
\affiliation{Department of Physics, University of Wisconsin-Madison, Madison, Wisconsin 53706, USA}

\date{\today}

\begin{abstract}
The ideal superconductor provides a pristine environment for the delicate states of a quantum computer: because there is an energy gap to excitations, there are no spurious modes with which the qubits can interact, causing irreversible decay of the quantum state. As a practical matter, however, there exists a high density of excitations out of the superconducting ground state even at ultralow temperature; these are known as quasiparticles. Observed quasiparticle densities are of order 1~$\mu$m$^{-3}$, tens of orders of magnitude larger than the equilibrium density expected from theory. Nonequilibrium quasiparticles extract energy from the qubit mode and induce discrete changes in qubit offset charge, a potential source of dephasing. Here we show that a dominant mechanism for quasiparticle poisoning in superconducting qubits is direct absorption of high-energy photons at the qubit junction. We use a Josephson junction-based photon source to controllably dose qubit circuits with millimeter-wave radiation, and we use an interferometric quantum gate sequence to reconstruct the charge parity on the qubit island. We find that the structure of the qubit itself acts as a resonant antenna for millimeter-wave radiation, providing an efficient path for photons to generate quasiparticle excitations. A deep understanding of this physics will pave the way to realization of next-generation superconducting qubits that are robust against quasiparticle poisoning and could enable a new class of quantum sensors for dark matter detection. %FOR SUBMISSION TO NATURE WE WILL NEED TO INCLUDE REFERENCES IN THE ABSTRACT -- RFM
\end{abstract}

\maketitle
In equilibrium, the ratio of thermally generated quasiparticles to Cooper pairs in a superconducting device is of order $10^{-50}$ at the millikelvin temperatures that are relevant for quantum computing and sensing applications.  This is due to the exponential suppression of  quasiparticle density with respect to $\Delta/T$, where $\Delta$ is the superconducting gap energy and $T$ is temperature \cite{Catelani11, Glazman2021}.  Experimentally, however, this ratio is found to be between $\sim10^{-6}$ and $10^{-10}$ \cite{Aumentado04,shaw2008kinetics,Martinis09,Vool14,Wang14,DeVisser2014}, more than 40 orders of magnitude larger than expected from the equilibrium calculation. Nonequilibrium quasiparticles limit the sensitivity of superconducting devices for charge sensing \cite{Aumentado04, Mannila2021}, metrology \cite{van2016single}, and astrophysical observation \cite{Day2003a, Echternach2018}. In the context of superconducting qubits, nonequilibrium quasiparticles represent a significant decoherence channel \cite{Martinis09, Catelani11,Vool14, Wang14, DeVisser2014, Serniak2018}. Recent experiments have demonstrated that quasiparticles liberated by particle impacts in the qubit substrate give rise to correlated relaxation errors in multiqubit arrays \cite{Vepsalainen2020, Wilen2021a, McEwen2021}. While such errors are especially damaging for quantum error correction, the particle impact rate is too low and the rate of removal of pair-breaking energy in the aftermath of an impact too high to account for the large baseline density of quasiparticles in superconducting quantum circuits.

\begin{figure*}[ht!]
\includegraphics[width=\columnwidth]{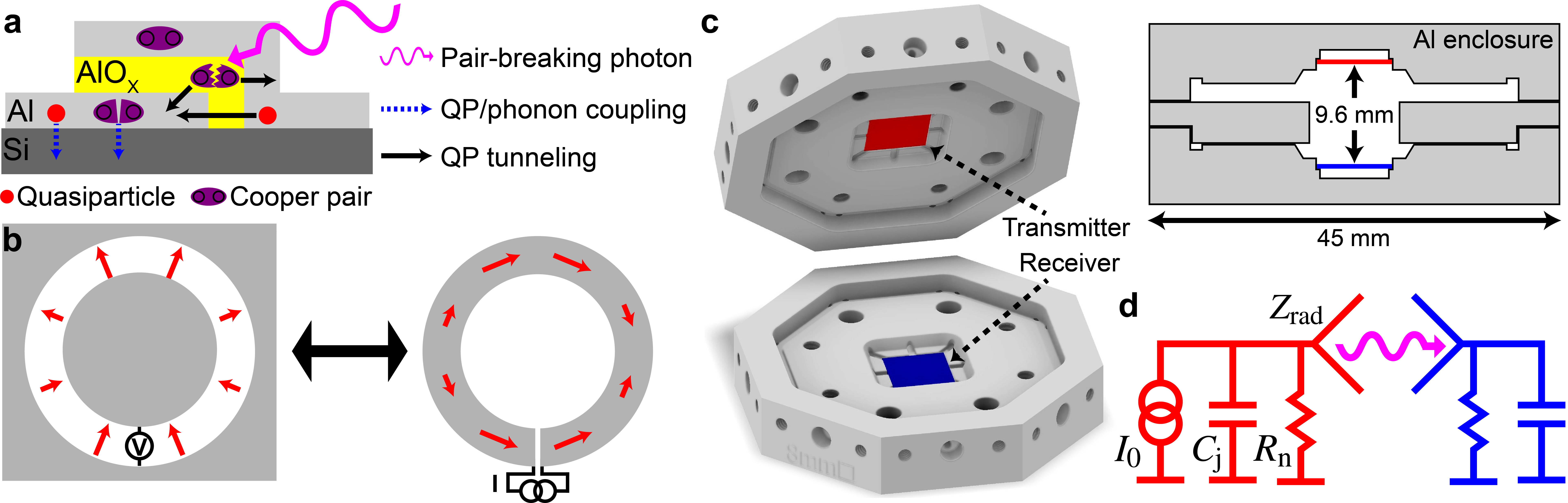}
\caption{\textbf{Photon-assisted quasiparticle poisoning mediated by spurious antenna modes of the qubit.} 
\textbf{a}, Quasiparticle generation and conversion processes. Pair-breaking radiation is absorbed at a Josephson junction. Photon absorption changes the charge parity of the qubit island and can induce a qubit state transition. Quasiparticles couple to phonons through scattering and recombination. 
\textbf{b}, The single-ended circular transmon (left) is the aperture dual of the resonant wire loop antenna (right). Red arrows show the amplitude and direction of electric fields. In the aperture antenna, signal is coupled via a high-impedance source at the voltage antinode, while in the dual wire antenna signal is coupled via a low-impedance source at the current antinode \cite{Rafferty2021a}.
\textbf{c}, Two-chip transmit/receive geometry used to probe the spectral response of the qubits to mm-wave radiation. Cutaway drawing to the right is a scale illustration of the aluminum sample enclosure.
\textbf{d}, Circuit diagram for the transmit/receive experiment, with mm-wave Josephson transmitter depicted in red and receiver qubit depicted in blue.
}
\label{fig:AntennaCartoon}
\end{figure*}

Another potential source of quasiparticles is the absorption of pair-breaking photons. It has been shown that improvements in filtering and shielding can lead to enhanced energy relaxation times for superconducting resonators \cite{Barends08} and qubits \cite{Corcoles2011a}. Recently, Houzet \textit{et al.} explained the observed ratio of parity switches to state transitions in a transmon qubit in terms of photon-assisted pair breaking at the Josephson junction \cite{Houzet19}; however, this work did not address the concrete physical mechanism by which the photon couples to the qubit junction. We recently put forth a model for the resonant absorption of pair-breaking photons by spurious antenna modes of the qubit \cite{Rafferty2021a}. The crucial insight is that the qubit structure itself exhibits a parasitic resonance at a frequency of order 100~GHz, set by the round-trip distance around the qubit island. This resonance is the aperture dual of the resonant wire loop antenna \cite{balanis2016antenna}. For typical qubit parameters, the qubit junction is well matched to free space impedance via this antenna mode, so that the qubit is an efficient absorber of pair-breaking radiation. Fig. \ref{fig:AntennaCartoon}a summarizes relevant quasiparticle generation and conversion mechanisms in superconducting qubits, while Fig. \ref{fig:AntennaCartoon}b depicts the dual mapping of aperture antenna to wire antenna. 

In this article, we describe the experimental validation of our model for the antenna coupling of qubits to pair-breaking radiation. The experiments involve two separate chips that are housed in a single enclosure; one chip incorporates voltage-biased Josephson junctions that act as transmitters of coherent mm-wave photons, while the second chip supports multiple superconducting qubits that act as receivers. We use a Ramsey-based interferometric gate sequence to monitor the charge-parity state of the qubits \cite{Riste2013}; resonant absorption of pair-breaking photons induces parity switches on the qubit which we detect with near unit fidelity. By scanning the voltage bias of the transmitter junction, we map out the spectral response of the qubits up to $\sim$500~GHz. We find that the detailed absorption spectrum of the qubits agrees well with the predictions of our model. 
In addition, we find that the baseline quasiparticle poisoning rate (in the absence of mm-wave injection) can be explained in terms of the resonant absorption of blackbody photons from higher temperature stages of the cryostat. Finally, we show that spurious transitions of the qubit out of the ground state are dominated by the resonant absorption of pair-breaking photons. This detailed understanding of the physical mechanism for quasiparticle poisoning will allow realization of new qubit designs that are robust against pair-breaking radiation; additionally, it could form the basis for a new class of quantum sensors based on the transduction of photons to quasiparticles followed by subsequent qubit-based detection. We note that a recent experimental study explains the scaling of quasiparticle poisoning rate with qubit size in terms of our model, without direct validation of the detailed spectral response of the qubit devices \cite{Pan2022}.

\begin{figure*}[ht!]
\includegraphics[width=\columnwidth]{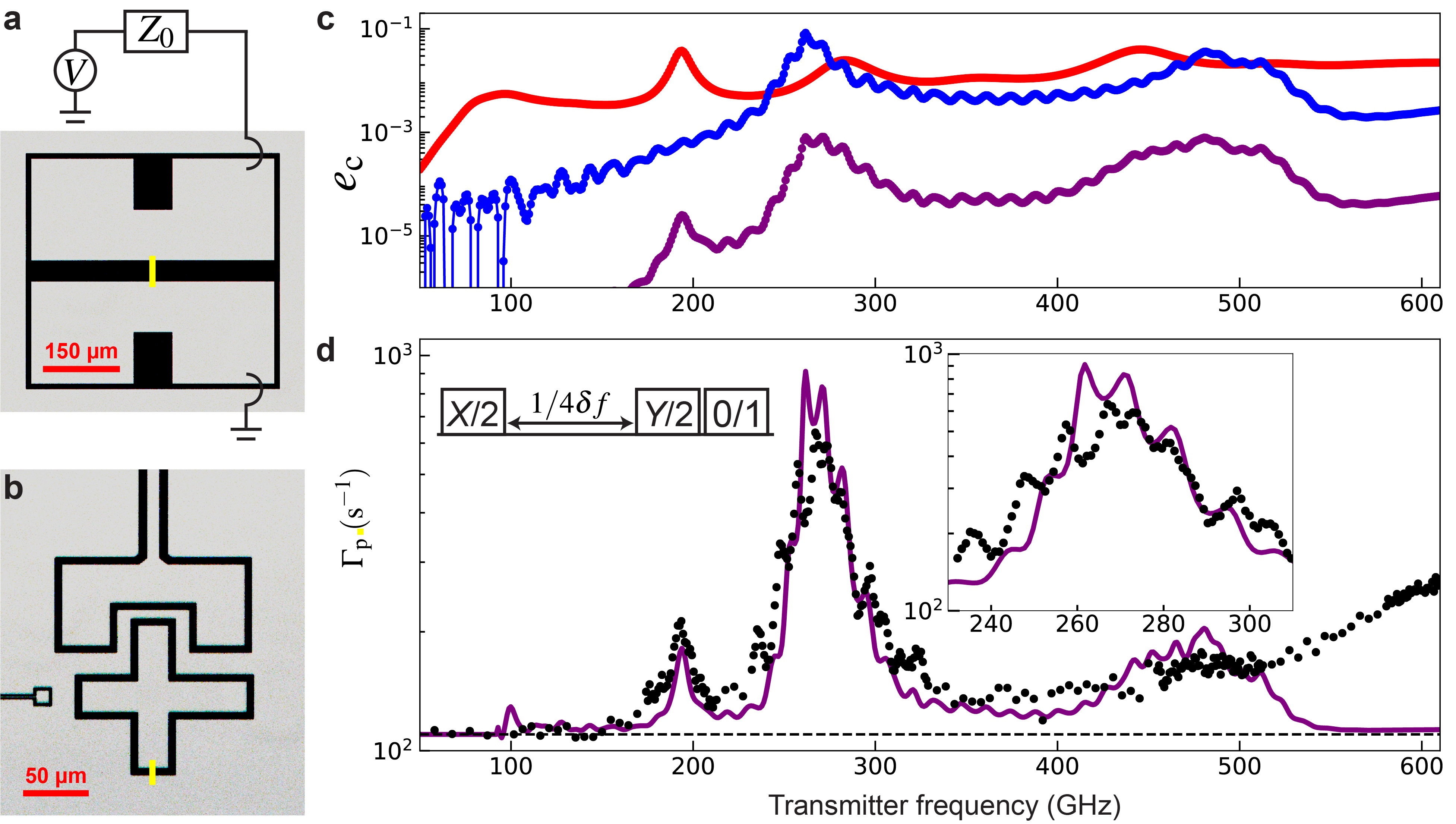}
\caption{\textbf{Spectral response of the Xmon qubit.} 
\textbf{a}, Optical micrograph of the Josephson transmitter. Voltage bias is provided by a high-bandwidth line with impedance $Z_0=50\,\Omega$. 
\textbf{b}, Optical micrograph of the Xmon receiver qubit. Local charge gate is shown at left and coupling capacitor to the readout resonator is shown at top. In both (a) and (b), junction leads are shown in yellow.
\textbf{c}, Frequency-dependent coupling efficiency calculated for the transmitter junction (red) and receiver qubit (blue). The purple trace, obtained as the product of these coupling efficiencies, represents the overall transfer efficiency from transmitter to receiver in the absence of losses and other nonidealities.
\textbf{d}, Quasiparticle poisoning rate as a function of transmitter frequency. Left inset shows the parity-sensitive Ramsey sequence. Black points are the measured poisoning rates; the black dashed line is the baseline rate $\Gamma_0 = 110 ~\rm s ^{-1}$ (in the absence of explicit photon injection); and the purple trace is the contribution from Josephson radiation calculated from the coupling efficiencies of (c), with an overall scaling of 0.07 to account for photon losses. The right inset shows a detailed view of the resonant features around 270~GHz. Oscillations in the spectral response of the qubit arise from the mutual coupling of the qubit antenna mode to a spurious slotline mode of the qubit readout resonator.
}
\label{fig:Xmon}
\end{figure*}

The experimental setup is shown in Fig. \ref{fig:AntennaCartoon}c. Two separate device chips are integrated in a single light-tight enclosure made from 6061 aluminum: the transmitter chip (red) is mounted face to face with the receiver chip (blue) with a separation of 9.6~mm. For all devices described here, the chip is nearly completely covered by a niobium groundplane, and all Josephson elements are realized as Al-AlO$_x$-Al tunnel junctions. We use a high-bandwidth 50~$\Omega$ line to bias the transmitter junction in the finite-voltage state; since the normal-state resistance of the junction is of order 10~k$\Omega$, the bias line represents a stiff voltage source. Bias of the transmitter junction at voltage $V$ induces coherent oscillations in the phase difference across the junction at the Josephson frequency $f_{\rm J} = V/\Phi_0$, where $\Phi_0 \equiv h/2e$ is the magnetic flux quantum; for a 1~mV bias, the Josephson frequency is 484~GHz. A circuit diagram of the experiment is shown in Fig. \ref{fig:AntennaCartoon}d. The voltage-biased junction can be modeled as a Norton equivalent current source $I_0$ in parallel with shunt admittance $Y_{\rm j} \equiv 1/Z_{\rm j} = 1/R_{\rm n} + j \omega C_{\rm j}$, where $I_0$ is the junction critical current, $R_{\rm n}$ is the junction normal-state resistance, and $C_{\rm j}$ is the self-capacitance of the junction. The Josephson oscillator acts as a source that drives the antenna formed by the junction pads embedded in the circuit groundplane, with radiation impedance $Z_{\rm rad}$. Following the analysis of \cite{Rafferty2021a}, the power radiated to free space is given by 
\begin{equation}\label{eq:P_rad}
    P_{\rm rad}=\frac{e_{\rm c, tr}}{8} I_0^2 R_{\rm n},
\end{equation}
where $I_0$ and $R_{\rm n}$ are the transmitter junction critical current and normal-state resistance, respectively, and where the coupling efficiency $e_{\rm c, tr}$ of the junction to free space is given by
\begin{equation}\label{eq:ec}
e_{\rm c, tr} = 1-\left|\frac{Z_{\rm rad} - Z_{\rm j}^*}{Z_{\rm rad} + Z_{\rm j}}\right|^2.
\end{equation}
Power emitted into the shared cavity will populate the enclosure with coherent mm-wave radiation; due to multiple reflections from the enclosure walls, we expect the distribution of energy in the enclosure to be isotropic with random polarization (see Supplement for additional details).

We now consider the effect of the radiation on the transmon qubits of the receiver chip. The devices were designed with ratio $E_{\rm J}/E_{\rm c}$ of Josephson energy to single-electron charging energy around 25, so that the qubit 01 transition frequency is weakly sensitive to offset charge on the qubit island, with peak-to-peak charge dispersion $2\delta f$ of order 1~MHz. 
This makes it possible to monitor quasiparticle poisoning events in real time by using a Ramsey gate sequence that maps charge parity to qubit state occupation (see Supplement). As acoustic coupling between the transmitter and receiver chips is negligible, the transfer of energy from chip to chip proceeds via the emission and absorption of photons. Photons with frequency $f > 2\Delta_{\rm Al}/h$~=~92~GHz couple to the Josephson junction of the receiver qubit with efficiency $e_{\rm c, rec}$ defined as in Eq.~\ref{eq:ec}. Photon absorption breaks a Cooper pair: two quasiparticles are generated, one on either side of the junction, resulting in a change in the charge parity of the qubit island. 

In a first series of experiments, we employ two nominally identical chips as transmitter and receiver. Witness junctions arrayed around the perimeter of the chip are used as mm-wave transmitters; the geometry is shown in Fig.~\ref{fig:Xmon}a. One pad of the transmitter is connected to the chip groundplane via a short wirebond, while the opposite pad is bonded to the 50~$\Omega$ bias line \footnote{Detailed modeling of the chip geometry using CST Microwave reveals that loading of the transmitter antenna with the wirebond and 50~$\Omega$ drive line only slightly modifies the radiation impedance of the structure.}. The chip incorporates six Xmon qubits, each with a local readout resonator coupled to a common feedline; the qubit geometry is shown in Fig.~\ref{fig:Xmon}b. As we vary the voltage bias on the transmitter junction and thus the frequency of the emitted Josephson radiation, we use the parity-sensitive Ramsey sequence to probe the charge parity of the receiver qubit at a sampling rate of 20~kHz. In the following, we present data on a single transmitter/receiver pair, although all three of the receiver qubits studied on this chip displayed a similar spectral response (see Supplement). All experiments were conducted at the 16~mK base temperature of a closed-cycle dilution refrigerator.

In Fig.~\ref{fig:Xmon}c, we show the calculated coupling efficiencies $e_{\rm c}$ for the transmitter (red) and receiver (blue) structures; here we assume a junction specific capacitance of $75$~fF/$\mu$m$^2$ for both chips. The purple trace, obtained as the product of the transmitter and receiver efficiencies, represents the frequency-dependent transfer function of the two-chip experiment in the absence of loss and additional nonidealities. In Fig.~\ref{fig:Xmon}d, we plot the measured parity rate $\Gamma_{\rm p}$ of the receiver qubit versus transmitter Josephson frequency (black points). We observe clear peaks in the spectral response of the receiver qubit at 190~GHz and 270~GHz, with $\Gamma_{\rm p}$ a factor of 2 and 6 times larger than the baseline rate of 110 s$^{-1}$, respectively. We ascribe these features to antenna resonances in our Josephson transmitter and receiver qubit, which provide enhanced transfer of Josephson energy between the two chips. The solid purple trace represents the expected parity rate calculated from the device coupling efficiencies. Detailed modeling of the two-chip experiment incorporates several effects, including quasiparticle-induced suppression of the critical current of the transmitter junction and randomization of the direction and polarization of the emitted radiation due to multiple reflections from the walls of the cavity enclosure; for a detailed discussion, see Supplement. The model clearly captures the dominant resonant features of the parity spectrum; however, to obtain good agreement between the measured and expected parity rates, we must scale the expected rates by a factor of 0.07. The discrepancy in the absolute rates could be due to dielectric losses or to enhanced loss of photons that are initially emitted into the silicon substrate of the transmitter, as the near-continuous superconducting Nb groundplane of the transmitter chip will act as a barrier to transport of photons into the shared space of the two-chip enclosure. 

In the inset of Fig.~\ref{fig:Xmon}d, we plot on an expanded scale the measured and calculated parity rates. We observe clear fine structure in both the measured and calculated spectra, with a modulation of the receiver response at a period of 11~GHz. We understand this modulation to be due to the mutual coupling between the receiver qubit and its local readout resonator, which involves a spurious $\lambda/2$ slotline mode at a frequency that is roughly twice that of the $\lambda/4$ coplanar waveguide resonance at 6.058~GHz. We take the excellent agreement between the measured and calculated fine structure of the parity spectra as clear validation of our antenna model for coupling of the qubit to pair-breaking radiation. 

For transmitter bias above 1~mV, corresponding to Josephson frequency above 500~GHz, we observe an upturn in the measured parity rate that is not captured by our modeling. We ascribe this feature to the exchange of energy between the chips due to the incoherent emission of photons from the recombination of quasiparticles in the leads of the transmitter junction; detailed modeling of this physics is the subject of ongoing work.

Two additional qubits on the receiver chip were examined during the same cooldown (see Supplement). Both of these devices displayed baseline parity rates and spectral responses within 10\% of those measured on the device described here, despite different separation and relative orientation with respect to the transmitter junction. This observation lends support to our assumption that the experiment can be modeled in terms of an isotropic distribution of radiation in the shared cavity with random polarization. 

\begin{figure}[t]
\includegraphics[width=\columnwidth]{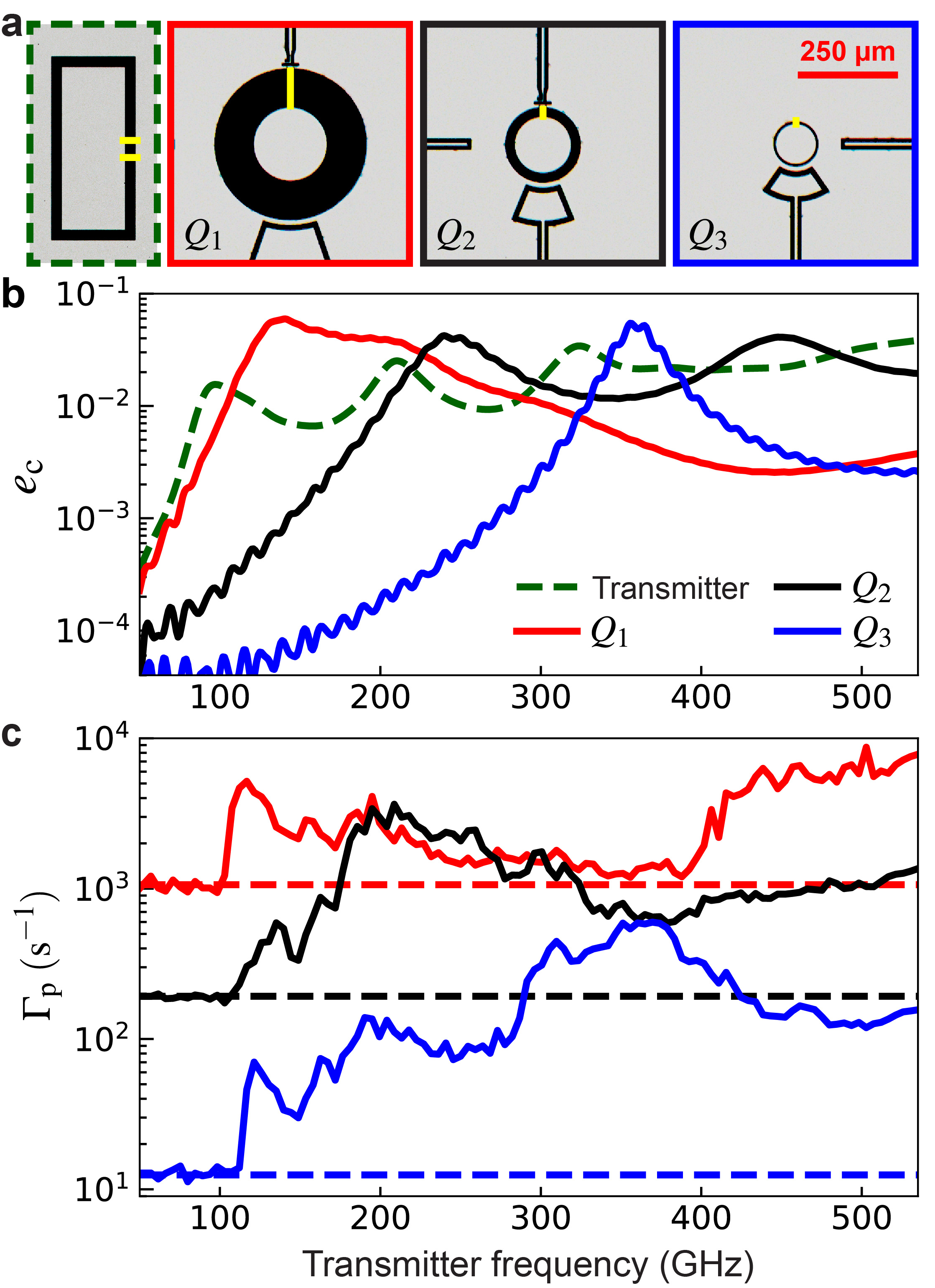}
\caption{\textbf{Dependence of resonant response and baseline parity rate on device scale.}
\textbf{a}, Optical micrographs of the rectangular transmitter device (green) and the large (red), intermediate (black), and small (blue) circmon qubits used for these experiments. Yellow traces indicate the junction leads.
\textbf{b}, Calculated coupling efficiencies of the transmitter and the three circmon receiver qubits.
\textbf{c}, Measured quasiparticle poisoning rates for the three qubits of as a function of transmitter frequency. Dashed lines indicate the baseline parity rates measured in the absence of photon injection.
}
\label{fig:Circmon}
\end{figure}
In a second series of experiments, we examined the resonant response of receiver qubits with circular island geometry spanning a range of sizes; the geometries of the transmitter junction and the three receiver qubits  ($Q_1$, $Q_{2}$, and $Q_{3}$) are shown in Fig.~\ref{fig:Circmon}a. The devices are designed with the same nominal charging energy $E_{\rm c}/h$~=~360~MHz and ratio $E_{\rm J}/E_{\rm c}$~=~28; however, the different island radii 90, 70, and 50 $\mu$m yield different dominant dipole antenna resonances at frequencies 130, 240, and 360~GHz, as confirmed by numerical modeling of the chip (Fig.~\ref{fig:Circmon}b).

With the Josephson radiator turned off, we first measure the baseline parity rates on the three devices (dashed lines in Fig.~\ref{fig:Circmon}c), finding $\Gamma_{0}(Q_{1})=1060$ s$^{-1}$, $\Gamma_{0}(Q_{2})=190$ s$^{-1}$, and  $\Gamma_{0}(Q_{3})=12.8$ s$^{-1}$. The two orders of magnitude discrepancy in the baseline parity rates across these devices indicates clearly that nonequilibrium quasiparticles are not uniformly distributed on the receiver chip, and that device geometry plays a critical role in the generation of quasiparticles. If we take the radiative environment of the qubit to be a blackbody at effective temperature $T$ and assume coupling of the qubit antenna to a single mode and polarization of the radiation field, we find a rate of absorption of pair-breaking photons given by
\begin{equation}
    \Gamma_0 = \int \frac{e_{\rm c}}{e^{h f /k_{\rm B}T}-1}\,\,df.
\end{equation}
From the measured parity rates on the three devices, we infer effective blackbody temperatures $T(Q_{1})=410$~mK, $T(Q_{2})=490$~mK, and $T(Q_{3})=460$~mK. We believe that the broadband pair-breaking photons giving rise to the observed parity jumps are not due to a single radiator at a physical temperature of 400-500~mK, but rather due to light leakage from higher temperature stages of the refrigerator (most likely via the coaxial wiring) that is insufficiently attenuated by the in-line Eccosorb filters. There is no reason to expect the spectrum of leakage radiation to follow that of an ideal blackbody; the discrepancy in the effective temperatures inferred for the three qubit antenna modes could reflect structure in the environmental spectrum. 

With the transmitter junction biased in the voltage state, we map out the resonant response of these three devices using the parity spectroscopy technique described above; the data are shown in Fig.~\ref{fig:Circmon}c. The complex resonant structure of the transmitter mode leads to rich structure in the resonant response of the three qubits; however, the measured parity rates are in qualitative agreement with our antenna model, with the resonant response shifting to higher frequency as the radius of the qubit island decreases.

\begin{figure}[t]
\includegraphics[width=\columnwidth]{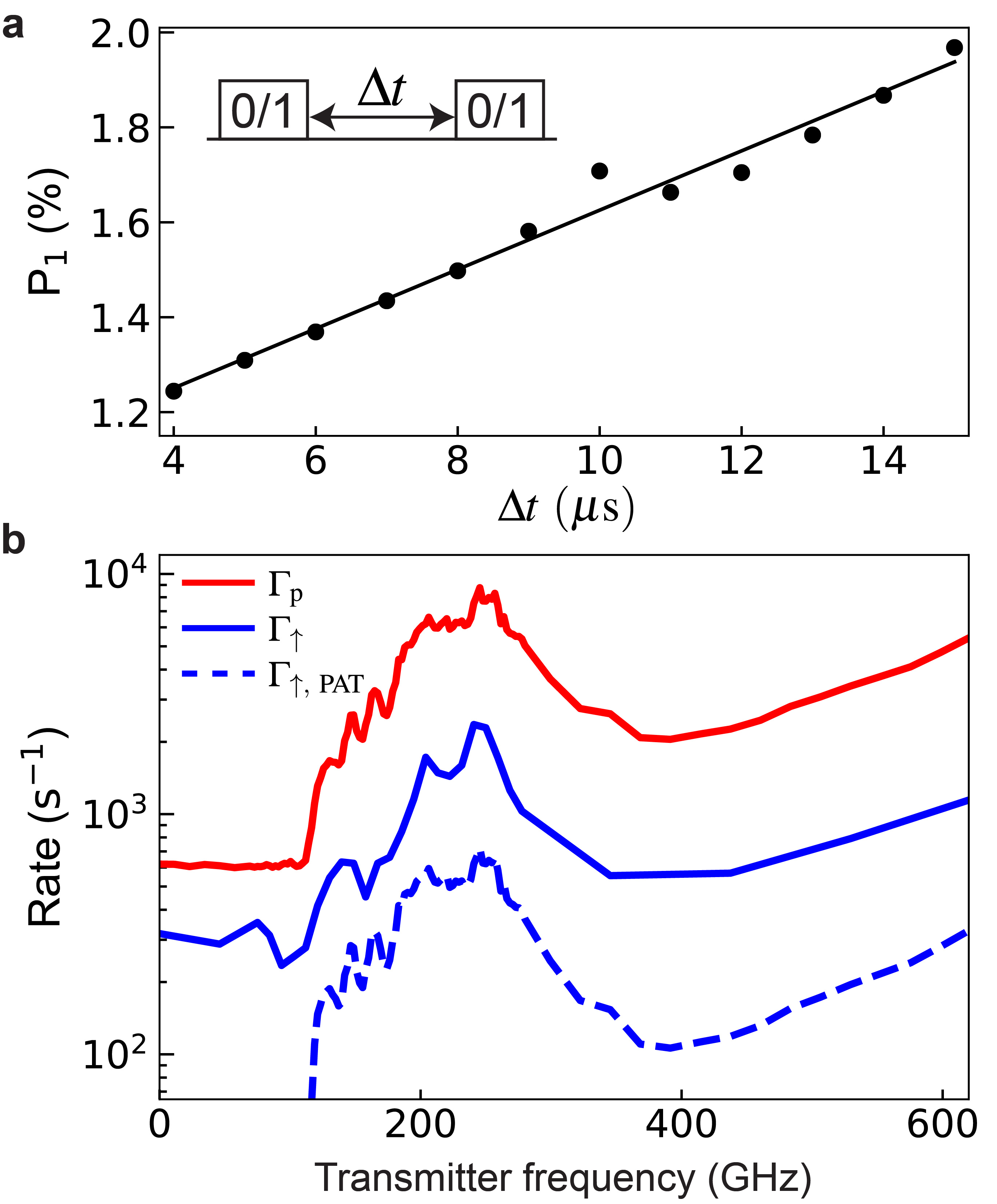}
\caption{\textbf{Photon-assisted parity switches and qubit transitions.} 
\textbf{a}, Representative measurement of qubit excitation rate. Here we use circmon qubit $Q_2$ as a testbed. Inset shows the $measure - idle - measure$ sequence. A linear fit (black line) is used to extract the upward transition rate.
\textbf{b}, Measured quasiparticle poisoning rate (red) and upward transition rate (blue solid trace) as a function of transmitter frequency. The dashed curve shows the predicted photon-assisted qubit transition rate calculated from the measured rate of parity switches, after \cite{Houzet19}. See Supplement for the details of the transmitter used in these experiments.
}
\label{fig:Up}
\end{figure}

Finally, using $Q_{2}$ as a testbed, we examine spurious transitions out of the qubit $\ket{0}$ state induced by the absorption of pair-breaking photons.
We measure the transition rate directly, using a $measure - idle - measure$ sequence (inset in Fig.~\ref{fig:Up}a), with the pre-measurement providing a high-fidelity initialization in the qubit $\ket{0}$ state. In Fig.~\ref{fig:Up}a we show representative data for the conditional probability $P_1 \equiv P(1|0)$ of finding the qubit in state $\ket{1}$ in the second measurement given that the initial measurement prepared state $\ket{0}$. A linear fit to the plot of $P_1$ versus idle time $\Delta t$ yields the upward transition rate $\Gamma_\uparrow$. In  Fig.~\ref{fig:Up}b, the solid traces show the measured parity rate $\Gamma_{\rm p}$ and upward transition rate $\Gamma_\uparrow$ versus Josephson frequency of the transmitter junction. We see that $\Gamma_{\rm p}$ and $\Gamma_\uparrow$ display a similar resonant response centered at a Josephson frequency around 240~GHz, where the transfer function from transmitter to receiver device is expected to peak. Moreover, the ratio $\Gamma_{\rm p}/\Gamma_\uparrow$ is roughly constant over the full frequency range, indicating that qubit transitions out of the ground state are dominated by resonant absorption of pair-breaking photons. Houzet \textit{et al.} have previously analyzed the rate of qubit transitions conditioned on the absorption of a pair-breaking photon \cite{Houzet19}. Their analysis predicts a contribution to $\Gamma_\uparrow$ given by the dashed blue trace in Fig.~\ref{fig:Up}b (see Supplement). However, the absorption of a pair-breaking photon will also generate a steady-state population of nonequilibrium quasiparticles that can tunnel across the qubit junction, inducing additional qubit transitions following the primary poisoning event. It is likely that this secondary poisoning process accounts for the enhanced rate  of upward transitions measured here.

In summary, we have used controlled irradiation of superconducting qubits with mm-wave photons derived from the ac Josephson effect to validate a model for photon-assisted quasiparticle poisoning through the spurious antenna modes of the transmon qubit. The observed baseline parity rates are well explained by absorption of broadband thermal photons from higher temperature stages of the cryostat. Additionally, the correlation between qubit state transitions and charge parity switches indicates that resonant absorption of pair-breaking photons is the dominant contributor to qubit initialization errors.

An understanding of the physical origin of quasiparticle poisoning will allow the development of improved qubit designs and measurement configurations that protect against absorption of pair-breaking radiation. At the same time, the resonant transduction of pair-breaking photons to quasiparticles followed by qubit-based parity detection could form the basis for a new class of quantum sensors; potential applications include high-resolution spectroscopy of the cosmic microwave background \cite{Wang2021} or detection of dark-matter axions \cite{Dixit2021a,Braine2020} or dark energy \cite{Ghosh2021} in the frequency range from 100~GHz to 1~THz, where established detection techniques are limited. 

Work at UW-Madison is supported by the Fermi National Accelerator
Laboratory, managed and operated by Fermi Research Alliance, LLC under Contract No. DE-AC02-07CH11359 with the U.S. Department of Energy, through the Office
of High Energy Physics QuantISED program. Work at Syracuse University is supported by the U.S. Government under ARO grant W911NF-18-1-0106. We thank E. S. Joseph and M. Wolfe for assistance with device imaging. This research was supported by an appointment to the Intelligence Community Postdoctoral Research Fellowship Program at University of Wisconsin-Madison, administered by Oak Ridge Institute for Science and Education through an interagency agreement between the U.S. Department of Energy and the Office of the Director of National Intelligence. Portions of this work were performed at the UW-Madison Wisconsin Centers for Nanoscale Technology, partially supported by the NSF through the University of Wisconsin Materials Research Science and Engineering Center (DMR-1720415).  Other portions were performed at the Cornell NanoScale Facility, a member of the National Nanotechnology Coordinated Infrastructure (NNCI), which is supported by the National Science Foundation under Grant No. NNCI-2025233.  

% \Liu{To Dave, let's try to split the ref after submission to arXiv.}
\bibliographystyle{naturemag}
\bibliography{RFM_refs_Feb20, AntennaModeBib, LocalBib, CL_nonstandard_ref}

\begin{thebibliography}{10}
\expandafter\ifx\csname url\endcsname\relax
  \def\url#1{\texttt{#1}}\fi
\expandafter\ifx\csname urlprefix\endcsname\relax\def\urlprefix{URL }\fi
\providecommand{\bibinfo}[2]{#2}
\providecommand{\eprint}[2][]{\url{#2}}

\bibitem{Catelani11}
\bibinfo{author}{Catelani, G.} \emph{et~al.}
\newblock \bibinfo{title}{Quasiparticle relaxation of superconducting qubits in
  the presence of flux}.
\newblock \emph{\bibinfo{journal}{Phys. Rev. Lett.}}
  \textbf{\bibinfo{volume}{106}}, \bibinfo{pages}{077002}
  (\bibinfo{year}{2011}).

\bibitem{Glazman2021}
\bibinfo{author}{Glazman, L.~I.} \& \bibinfo{author}{Catelani, G.}
\newblock \bibinfo{title}{{Bogoliubov quasiparticles in superconducting qubits
  Superconductivity in an isolated metallic island Electron pairing and
  condensate}}.
\newblock \emph{\bibinfo{journal}{SciPost Phys. Lect.Notes}}
  \textbf{\bibinfo{volume}{31}}, \bibinfo{pages}{1--40} (\bibinfo{year}{2021}).

\bibitem{Aumentado04}
\bibinfo{author}{Aumentado, J.}, \bibinfo{author}{Keller, M.~W.},
  \bibinfo{author}{Martinis, J.~M.} \& \bibinfo{author}{Devoret, M.~H.}
\newblock \bibinfo{title}{Nonequilibrium quasiparticles and $2e$ periodicity in
  single-{C}ooper-pair transistors}.
\newblock \emph{\bibinfo{journal}{Phys. Rev. Lett.}}
  \textbf{\bibinfo{volume}{92}}, \bibinfo{pages}{066802}
  (\bibinfo{year}{2004}).

\bibitem{shaw2008kinetics}
\bibinfo{author}{Shaw, M.}, \bibinfo{author}{Lutchyn, R.},
  \bibinfo{author}{Delsing, P.} \& \bibinfo{author}{Echternach, P.}
\newblock \bibinfo{title}{Kinetics of nonequilibrium quasiparticle tunneling in
  superconducting charge qubits}.
\newblock \emph{\bibinfo{journal}{Phys. Rev. B}} \textbf{\bibinfo{volume}{78}},
  \bibinfo{pages}{024503} (\bibinfo{year}{2008}).

\bibitem{Martinis09}
\bibinfo{author}{Martinis, J.~M.}, \bibinfo{author}{Ansmann, M.} \&
  \bibinfo{author}{Aumentado, J.}
\newblock \bibinfo{title}{Energy decay in superconducting {J}osephson-junction
  qubits from nonequilibrium quasiparticle excitations}.
\newblock \emph{\bibinfo{journal}{Phys. Rev. Lett.}}
  \textbf{\bibinfo{volume}{103}}, \bibinfo{pages}{097002}
  (\bibinfo{year}{2009}).

\bibitem{Vool14}
\bibinfo{author}{Vool, U.} \emph{et~al.}
\newblock \bibinfo{title}{Non-{P}oissonian quantum jumps of a fluxonium qubit
  due to quasiparticle excitations}.
\newblock \emph{\bibinfo{journal}{Phys. Rev. Lett.}}
  \textbf{\bibinfo{volume}{113}}, \bibinfo{pages}{247001}
  (\bibinfo{year}{2014}).

\bibitem{Wang14}
\bibinfo{author}{Wang, C.} \emph{et~al.}
\newblock \bibinfo{title}{Measurement and control of quasiparticle dynamics in
  a superconducting qubit}.
\newblock \emph{\bibinfo{journal}{Nat. Commun.}} \textbf{\bibinfo{volume}{5}},
  \bibinfo{pages}{5836} (\bibinfo{year}{2014}).

\bibitem{DeVisser2014}
\bibinfo{author}{{De Visser}, P.~J.}, \bibinfo{author}{Baselmans, J.~J.},
  \bibinfo{author}{Bueno, J.}, \bibinfo{author}{Llombart, N.} \&
  \bibinfo{author}{Klapwijk, T.~M.}
\newblock \bibinfo{title}{{Fluctuations in the electron system of a
  superconductor exposed to a photon flux}}.
\newblock \emph{\bibinfo{journal}{Nat. Commun.}} \textbf{\bibinfo{volume}{5}},
  \bibinfo{pages}{3130} (\bibinfo{year}{2014}).

\bibitem{Mannila2021}
\bibinfo{author}{Mannila, E.~T.} \emph{et~al.}
\newblock \bibinfo{title}{{A superconductor free of quasiparticles for
  seconds}}.
\newblock \emph{\bibinfo{journal}{Nat. Phys.}} \textbf{\bibinfo{volume}{18}},
  \bibinfo{pages}{145--148} (\bibinfo{year}{2022}).

\bibitem{van2016single}
\bibinfo{author}{Van~Zanten, D.} \emph{et~al.}
\newblock \bibinfo{title}{Single quantum level electron turnstile}.
\newblock \emph{\bibinfo{journal}{Phys. Rev. Lett.}}
  \textbf{\bibinfo{volume}{116}}, \bibinfo{pages}{166801}
  (\bibinfo{year}{2016}).

\bibitem{Day2003a}
\bibinfo{author}{Day, P.}, \bibinfo{author}{LeDuc, H.}, \bibinfo{author}{Mazin,
  B.}, \bibinfo{author}{Anastasios, V.} \& \bibinfo{author}{Zmuidzinas, J.}
\newblock \bibinfo{title}{{A broadband superconducting detector suitable for
  use in large arrays}}.
\newblock \emph{\bibinfo{journal}{Nature}} \textbf{\bibinfo{volume}{425}},
  \bibinfo{pages}{817--821} (\bibinfo{year}{2003}).

\bibitem{Echternach2018}
\bibinfo{author}{Echternach, P.~M.}, \bibinfo{author}{Pepper, B.~J.},
  \bibinfo{author}{Reck, T.} \& \bibinfo{author}{Bradford, C.~M.}
\newblock \bibinfo{title}{{Single photon detection of 1.5 THz radiation with
  the quantum capacitance detector}}.
\newblock \emph{\bibinfo{journal}{Nat. Astron.}} \textbf{\bibinfo{volume}{2}},
  \bibinfo{pages}{90--97} (\bibinfo{year}{2018}).

\bibitem{Serniak2018}
\bibinfo{author}{Serniak, K.} \emph{et~al.}
\newblock \bibinfo{title}{{Hot Nonequilibrium Quasiparticles in Transmon
  Qubits}}.
\newblock \emph{\bibinfo{journal}{Phys. Rev. Lett.}}
  \textbf{\bibinfo{volume}{121}}, \bibinfo{pages}{157701}
  (\bibinfo{year}{2018}).

\bibitem{Vepsalainen2020}
\bibinfo{author}{Veps{\"{a}}l{\"{a}}inen, A.~P.} \emph{et~al.}
\newblock \bibinfo{title}{{Impact of ionizing radiation on superconducting
  qubit coherence}}.
\newblock \emph{\bibinfo{journal}{Nature}} \textbf{\bibinfo{volume}{584}},
  \bibinfo{pages}{551--556} (\bibinfo{year}{2020}).

\bibitem{Wilen2021a}
\bibinfo{author}{Wilen, C.~D.} \emph{et~al.}
\newblock \bibinfo{title}{{Correlated charge noise and relaxation errors in
  superconducting qubits}}.
\newblock \emph{\bibinfo{journal}{Nature}} \textbf{\bibinfo{volume}{594}},
  \bibinfo{pages}{369--373} (\bibinfo{year}{2021}).

\bibitem{McEwen2021}
\bibinfo{author}{McEwen, M.} \emph{et~al.}
\newblock \bibinfo{title}{{Resolving catastrophic error bursts from cosmic rays
  in large arrays of superconducting qubits}}.
\newblock \emph{\bibinfo{journal}{Nat. Phys.}} \textbf{\bibinfo{volume}{18}},
  \bibinfo{pages}{107--111} (\bibinfo{year}{2021}).

\bibitem{Barends08}
\bibinfo{author}{Barends, R.} \emph{et~al.}
\newblock \bibinfo{title}{Contribution of dielectrics to frequency and noise of
  {NbTiN} superconducting resonators}.
\newblock \emph{\bibinfo{journal}{Appl. Phys. Lett.}}
  \textbf{\bibinfo{volume}{92}}, \bibinfo{pages}{223502}
  (\bibinfo{year}{2008}).

\bibitem{Corcoles2011a}
\bibinfo{author}{C{\'{o}}rcoles, A.~D.} \emph{et~al.}
\newblock \bibinfo{title}{{Protecting superconducting qubits from radiation}}.
\newblock \emph{\bibinfo{journal}{Appl. Phys. Lett.}}
  \textbf{\bibinfo{volume}{99}}, \bibinfo{pages}{181906}
  (\bibinfo{year}{2011}).

\bibitem{Houzet19}
\bibinfo{author}{Houzet, M.}, \bibinfo{author}{Serniak, K.},
  \bibinfo{author}{Catelani, G.}, \bibinfo{author}{Devoret, M.~H.} \&
  \bibinfo{author}{Glazman, L.~I.}
\newblock \bibinfo{title}{Photon-assisted charge-parity jumps in a
  superconducting qubit}.
\newblock \emph{\bibinfo{journal}{Phys. Rev. Lett.}}
  \textbf{\bibinfo{volume}{123}}, \bibinfo{pages}{107704}
  (\bibinfo{year}{2019}).

\bibitem{Rafferty2021a}
\bibinfo{author}{Rafferty, O.} \emph{et~al.}
\newblock \bibinfo{title}{{Spurious Antenna Modes of the Transmon Qubit.}}
  \bibinfo{pages}{arXiv:2103.06803} (\bibinfo{year}{2021}).

\bibitem{balanis2016antenna}
\bibinfo{author}{Balanis, C.~A.}
\newblock \emph{\bibinfo{title}{Antenna Theory: Analysis and Design}}
  (\bibinfo{publisher}{Wiley}, \bibinfo{year}{2016}).

\bibitem{Riste2013}
\bibinfo{author}{Rist{\`e}, D.} \emph{et~al.}
\newblock \bibinfo{title}{Millisecond charge-parity fluctuations and induced
  decoherence in a superconducting transmon qubit}.
\newblock \emph{\bibinfo{journal}{Nat. Commun.}} \textbf{\bibinfo{volume}{4}},
  \bibinfo{pages}{1913} (\bibinfo{year}{2013}).

\bibitem{Pan2022}
\bibinfo{author}{Pan, X.} \emph{et~al.}
\newblock \bibinfo{title}{{Engineering superconducting qubits to reduce
  quasiparticles and charge noise.}} \bibinfo{pages}{arXiv:2202.01435}
  (\bibinfo{year}{2022}).

\bibitem{Note1}
\bibinfo{note}{Detailed modeling of the chip geometry using CST Microwave
  reveals that loading of the transmitter antenna with the wirebond and
  50~$\Omega $ drive line only slightly modifies the radiation impedance of the
  structure.}

\bibitem{Wang2021}
\bibinfo{author}{Wang, Z.} \emph{et~al.}
\newblock \bibinfo{title}{{Quantum Microwave Radiometry with a Superconducting
  Qubit}}.
\newblock \emph{\bibinfo{journal}{Phys. Rev. Lett.}}
  \textbf{\bibinfo{volume}{126}}, \bibinfo{pages}{180501}
  (\bibinfo{year}{2021}).

\bibitem{Dixit2021a}
\bibinfo{author}{Dixit, A.~V.} \emph{et~al.}
\newblock \bibinfo{title}{{Searching for Dark Matter with a Superconducting
  Qubit}}.
\newblock \emph{\bibinfo{journal}{Phys. Rev. Lett.}}
  \textbf{\bibinfo{volume}{126}}, \bibinfo{pages}{141302}, \eprint{2008.12231}
  (\bibinfo{year}{2021}).

\bibitem{Braine2020}
\bibinfo{author}{Braine, T.} \emph{et~al.}
\newblock \bibinfo{title}{{Extended Search for the Invisible Axion with the
  Axion Dark Matter Experiment}}.
\newblock \emph{\bibinfo{journal}{Phys. Rev. Lett.}}
  \textbf{\bibinfo{volume}{124}}, \bibinfo{pages}{101303}
  (\bibinfo{year}{2020}).

\bibitem{Ghosh2021}
\bibinfo{author}{Ghosh, S.}, \bibinfo{author}{Ruddy, E.~P.},
  \bibinfo{author}{Jewell, M.~J.}, \bibinfo{author}{Leder, A.~F.} \&
  \bibinfo{author}{Maruyama, R.~H.}
\newblock \bibinfo{title}{{Searching for dark photons with existing haloscope
  data}}.
\newblock \emph{\bibinfo{journal}{Phys. Rev. D}}
  \textbf{\bibinfo{volume}{104}}, \bibinfo{pages}{92016}
  (\bibinfo{year}{2021}).

\bibitem{Dolan77}
\bibinfo{author}{Dolan, G.~J.}
\newblock \bibinfo{title}{Offset masks for lift-off photoprocessing}.
\newblock \emph{\bibinfo{journal}{Appl. Phys. Lett.}}
  \textbf{\bibinfo{volume}{31}}, \bibinfo{pages}{337--339}
  (\bibinfo{year}{1977}).

\bibitem{Note2}
\bibinfo{note}{CST Studio Suite, www.3ds.com.}

\bibitem{Lenander11}
\bibinfo{author}{Lenander, M.} \emph{et~al.}
\newblock \bibinfo{title}{Measurement of energy decay in superconducting qubits
  from nonequilibrium quasiparticles}.
\newblock \emph{\bibinfo{journal}{Phys. Rev. B}} \textbf{\bibinfo{volume}{84}},
  \bibinfo{pages}{024501} (\bibinfo{year}{2011}).

\bibitem{Clark1970}
\bibinfo{author}{Clark, A.~F.}, \bibinfo{author}{Childs, G.~E.} \&
  \bibinfo{author}{Wallace, G.~H.}
\newblock \bibinfo{title}{{Electrical resistivity of some engineering alloys at
  low temperatures}}.
\newblock \emph{\bibinfo{journal}{Cryogenics (Guildf).}}
  \textbf{\bibinfo{volume}{10}}, \bibinfo{pages}{295--305}
  (\bibinfo{year}{1970}).

\bibitem{Ambegaokar82}
\bibinfo{author}{Ambegaokar, V.}, \bibinfo{author}{Eckern, U.} \&
  \bibinfo{author}{Sch\"on, G.}
\newblock \bibinfo{title}{Quantum dynamics of tunneling between
  superconductors}.
\newblock \emph{\bibinfo{journal}{Phys. Rev. Lett.}}
  \textbf{\bibinfo{volume}{48}}, \bibinfo{pages}{1745} (\bibinfo{year}{1982}).

\bibitem{Kaplan}
\bibinfo{author}{Kaplan, S.~B.} \emph{et~al.}
\newblock \bibinfo{title}{{Quasiparticle and phonon lifetimes in
  superconductors}}.
\newblock \emph{\bibinfo{journal}{Phys. Rev. B}} \textbf{\bibinfo{volume}{14}},
  \bibinfo{pages}{4854--4873} (\bibinfo{year}{1976}).

\end{thebibliography}

\newpage
\clearpage

\section*{SUPPLEMENT}
\beginsupplement

\subsection{Device Fabrication}
The devices were fabricated in a single-layer process on a high-resistivity silicon substrate ($>$10 k$\Omega$-cm) with 100 crystal orientation. Following a strip of the native SiO$_x$ in dilute (2\%) hydrofluoric acid for 1 minute, we sputter a 100-nm thick Nb film at a rate of 50~nm/minute. (In the case of the Xmon qubits, the Nb film thickness is 70~nm). We use an i-line stepper to define the qubit islands, resonators, and control wiring. Next, we etch the Nb using Cl$_2$/BCl$_3$ chemistry in an inductively coupled plasma reactive ion etch tool. 

The qubit and transmitter junctions are formed in a Dolan bridge process \cite{Dolan77} with an MMA/PMMA resist bilayer that is patterned with a 100-keV electron-beam writer. Double-angle evaporation of the Al-AlO$_x$-Al stack is performed in an electron-beam evaporation tool with base pressure $2\times10^{-8}$~Torr; prior to junction growth, we perform an \textit{in situ} ion mill clean of the sample to ensure good metallic contact to the Nb base metal layer.

\subsection{Experiment Setup}\label{sec:Wiring}
\begin{figure*}[t]
\includegraphics[width=\textwidth]{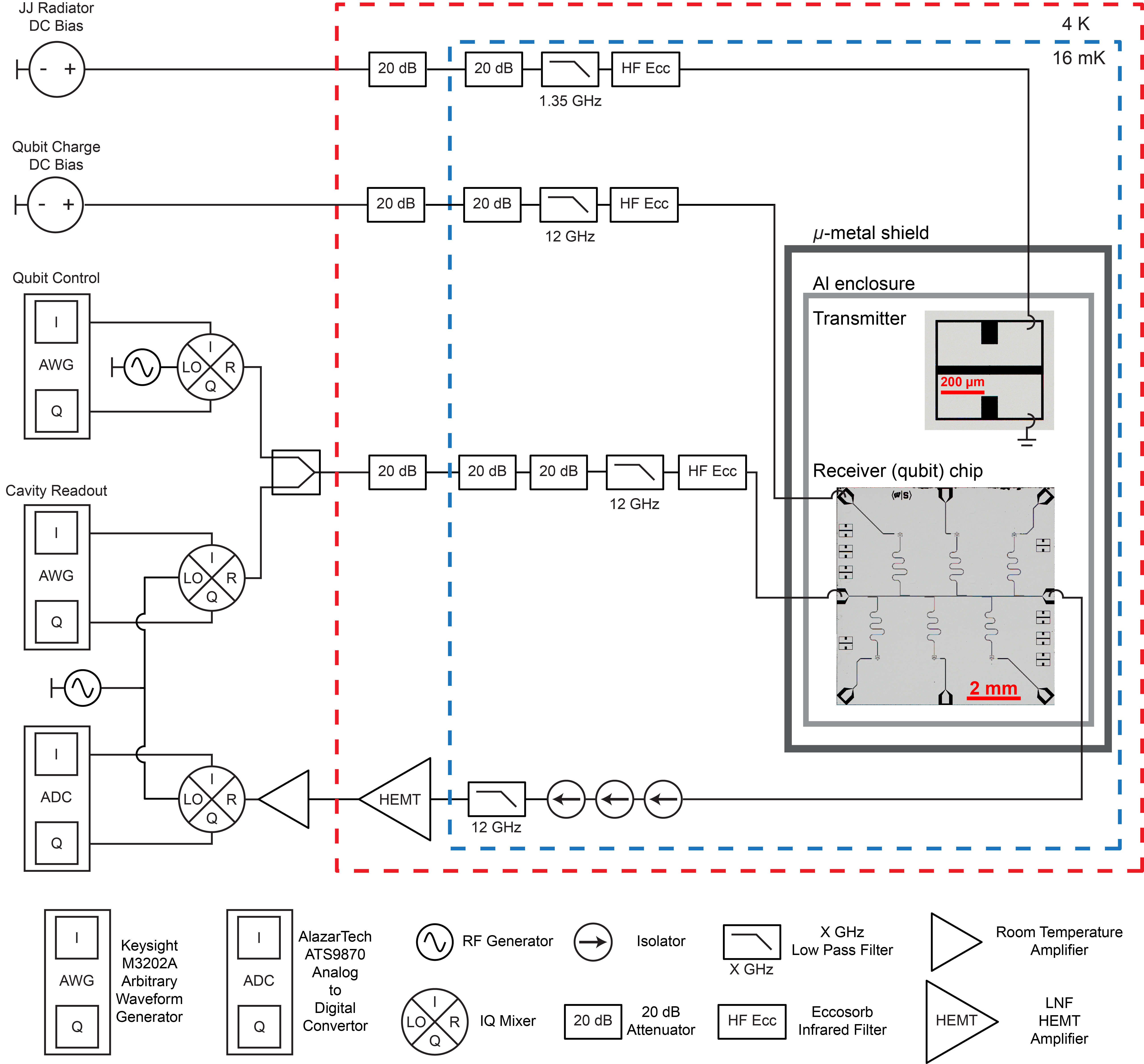}
\caption{\textbf{Wiring diagram of the experiments.}}
\label{fig:wiring_diagram} 
\end{figure*}

The experimental wiring configuration is shown in Fig.~\ref{fig:wiring_diagram}.

\subsection{Qubit Parameters}
In Table \ref{tab:qparams}, we list the measured and extracted parameters of the qubits used in these experiments.

\begin{table*}[t]
\caption{\textbf{Parameters of qubits used in the experiments.} Qubit frequency $f_{01}$ and peak-to-peak charge dispersion $2\delta f$ are measured by qubit spectroscopy. Ratio $E_{\rm J}/E_{\rm c}$ of Josephson energy to charging energy ratio is calculated from the measured qubit transition frequency and charge dispersion. The junction normal state resistance $R_{\rm n}$ is calculated from the extracted Josephson energy. The junction self-capacitance is obtained from the junction area measured from SEM and a nominal junction specific capacitance of 75~fF$/\mu$m$^2$. The frequency $f_{\rm ant}$ of the fundamental antenna resonance of the qubit is determined from the qubit antenna impedance calculated using a finite element solver and from the frequency-dependent junction impedance. The baseline quasiparticle poisoning rate $\Gamma_{0}$ is measured in the absence of explicit photon injection.}

\begin{tabular}{ |c||c|c|c|c|c|c|c| }
 \hline
 Xmon & $f_{01}$ (GHz) & $2\delta f$ (MHz)& $E_{\rm J}/E_{\rm c}$  & $R_{\rm n}$ (k$\Omega$) & $C_{\rm j}$ (fF)& $f_{\rm ant}$ (GHz)& $\Gamma_{0}$ (s$^{-1}$)\\
 \hline
   $Q_{1}$ & 4.828 & 4.6 & 22 & 17.1 & 4.6 &  280 & 85\\
 \hline
   $Q_{2}$ & 4.782 & 2.7 & 24 & 16.6 & 4.6 & 270 & 110\\
 \hline
  $Q_{3}$ & 4.829 & 2.1 & 25 & 16.1 & 4.6 &  270 & 103\\
 \hline
 \hline
 Circmon & $f_{01}$ (GHz) & $2\delta f$ (MHz)& $E_{\rm J}/E_{\rm c}$  & $R_{\rm n}$ (k$\Omega$) & $C_{\rm j}$ (fF)& $f_{\rm ant}$ (GHz)& $\Gamma_{0}$ (s$^{-1}$)\\
 \hline
   $Q_{1}$ & 4.595 & 1.5 & 26 & 16.6 & 3.5 &  130 & 1060\\
 \hline
   $Q_{2}$ & 4.949 & 1.1 & 28 & 15.0 & 3.2 & 240 & 190\\
 \hline
  $Q_{3}$ & 4.443 & 0.8 & 29 & 13.2 & 3.4 &  360 & 12.8\\
 \hline
\end{tabular}
\label{tab:qparams}
\end{table*}

\subsection{Charge Parity Measurement}\label{sec:PSD}
\begin{figure}[b!]
\includegraphics[width=\columnwidth]{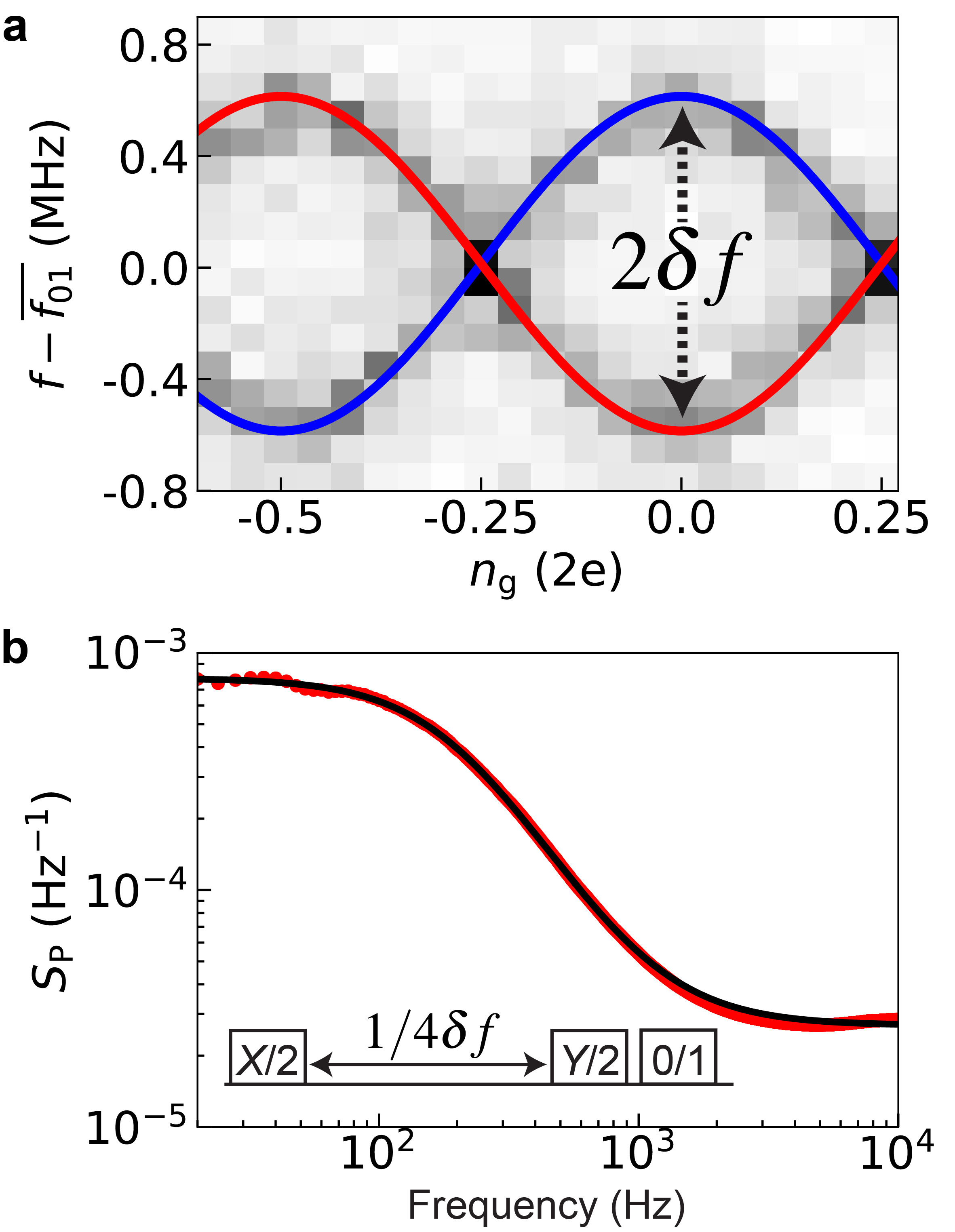}
\caption{\textbf{Charge parity measurement.} 
\textbf{a}, Spectroscopy of the charge-sensitive transmon versus applied offset charge $n_{g}$. Solid traces correspond to even and odd charge parity bands. Here we have maximum charge dispersion $2\delta f=$1.1 MHz. 
\textbf{b}, Power spectral density of charge-parity switches (red dots). The Lorentzian fit (black trace) corresponds to a parity switching rate $\Gamma_{\rm p}= 612 ~\rm s^{-1}$. Inset shows the Ramsey sequence used to map charge parity to qubit population.
}
\label{fig:S_chargeparity}
\end{figure}
We use the weakly charge-sensitive transmon to probe charge parity fluctuations. The devices are designed with $E_{\rm J}/E_{\rm c}= 20 - 30$, where $E_{\rm J}$ is the Josephson energy of the qubit junction and $E_{\rm c}$ is the single-electron charging energy of the qubit island, yielding a peak-to-peak charge dispersion of the qubit 01 transition around 1~MHz. In the following, we present data from circmon qubit $Q_{2}$ (described in Figs. \ref{fig:Circmon} and \ref{fig:Up} of the main text) as an example. For this device, $E_{\rm J}/E_{\rm c}= 28$.

In Fig.~\ref{fig:S_chargeparity}a, we show qubit spectroscopy versus applied gate charge. We observe two clear charge parity bands that oscillate sinusoidally with gate charge, with a maximum dispersion $2\delta f = 1.1$~MHz. To probe charge parity, we perform a series of charge-sensitive Ramsey scans at different applied charge biases in order to determine the offset charge on the qubit island; we then dynamically set the qubit charge bias to the point of maximum charge dispersion. We next apply a parity-sensitive $X/2 - idle - Y/2$ Ramsey sequence to map charge parity to qubit population (Fig. \ref{fig:S_chargeparity}b inset). We repeat the parity measurement sequence 5000 times with a duty cycle $\Delta t = 50~\mu$s to obtain a time series of charge parity. We repeat this experiment (including charge reset) 9 times, and from the separate time series we compute an average power spectral density $S_{\rm P}$ of charge parity fluctuations; representative data is shown in Fig.~\ref{fig:S_chargeparity}b. The power spectrum is fit using the Lorentzian form:
\begin{equation}
    S_{\rm P}(f)=\frac{4F^2\Gamma_{\rm p}}{(2\Gamma_{\rm p})^2+(2\pi f)^2} + (1-F^2)\Delta t,
\end{equation}
 where $F$ is the sequence mapping fidelity. From the fit, we extract the average parity switching rate $\Gamma_{\rm p}$.

\subsection{Analysis of Transmit/Receive Experiment}\label{sec:Analysis of TR}
Here we analyze the efficiency with which our transmitter and receiver devices are coupled to free space, and we put forth a simple model for estimating the circulating power in the shared cavity of the two chips. The analysis is focused on the transmit/receive experiment described in Fig.~\ref{fig:Xmon} of the main text. 

\subsubsection{Coupling Efficiency}
\begin{figure}[b!]
\includegraphics[width=\columnwidth]{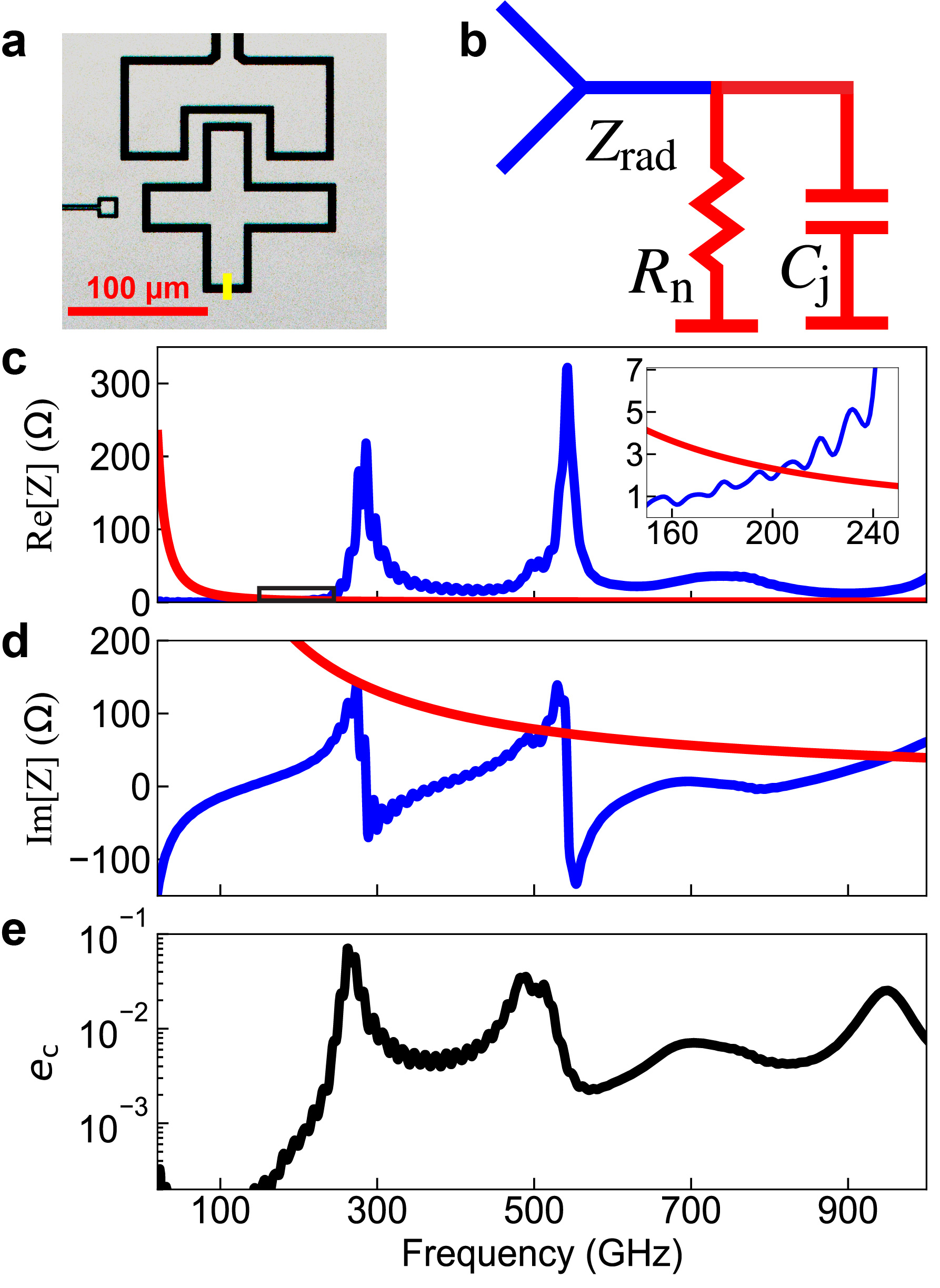}
\caption{
\textbf{Power match to the qubit antenna mode.} 
\textbf{a}, Single-ended Xmon from Fig.~\ref{fig:Xmon}b of the main text. The $X$-shaped qubit island is connected to ground by a Josephson junction (junction leads indicated in yellow). 
\textbf{b}, Equivalent circuit diagram for qubit junction (modeled as a parallel $RC$ circuit, red) embedded in an aperture antenna (blue). 
\textbf{c}, Real and \textbf{d}, imaginary parts of the conjugate of the junction impedance $Z_{\rm j}^*$ (red) and antenna impedance $Z_{\rm rad}$ (blue). The inset in \textbf{c} shows an expanded view of the real part of the impedance from 160 to 240~GHz.
\textbf{e}, Frequency-dependent coupling efficiency $e_{\rm c}$.
}
\label{fig:S_AntennaImpedance}
\end{figure}
The single-ended planar qubit can be viewed as an aperture antenna loaded with a Josephson junction; see Fig.~\ref{fig:S_AntennaImpedance}a. The junction is modeled as the parallel combination of tunnel resistance $R_{\rm n}$ and junction self-capacitance $C_{\rm j}$; see Fig.~\ref{fig:S_AntennaImpedance}b. The frequency-dependent impedance of the junction is given by
\begin{equation}
    Z_{\rm j} = \frac{1- j\omega\tau}{1+\omega^2\tau^2}R_{\rm n},
\end{equation}
where $\tau \equiv R_{\rm n}C_{\rm j}$. The impedance of the aperture antenna $Z_{\rm rad}$ is obtained from a finite element simulation \footnote{CST Studio Suite, www.3ds.com.}. Here we use the design of $Q_{2}$ from the Xmon experiment as an example. In Fig.~\ref{fig:S_AntennaImpedance}c-d, we plot the real and imaginary parts of $Z_{\rm rad}$ and $Z_{\rm j}^{*}$. The coupling efficiency $e_c$ is given as
\begin{equation}
    e_{\rm c} = 1 - \left|\frac{Z_{\rm rad}-Z_{\rm j}^*}{Z_{\rm rad}+Z_{\rm j}}\right|^2.
\end{equation}
Optimal power transfer is achieved when the conjugate matching condition is satisfied: $Z_{\rm rad}=Z_{j}^*$.
The coupling efficiency of $Q_{2}$ is shown in Fig.~\ref{fig:S_AntennaImpedance}e.
For our transmit/receive experiments, the coupling efficiency of the transmitter junction $e_{\rm c,tr}$ is calculated in the same way.

\subsubsection{Radiation from the Transmitter}

The power emitted from the transmitter junction due to Josephson oscillations is given by
\begin{equation}
    P_{\rm rad} =P_{\rm in, tr}e_{\rm c, tr},
\label{eq:P_rad_sup}
\end{equation}
where $e_{\rm c, tr}$ is calculated as in Eq.~\ref{eq:ec} from the main text, and where
\begin{equation}
    P_{\rm in, tr} =\frac{1}{8}I_0^2R_{\rm n} 
\end{equation}
is the maximum available power. Here, one factor of 1/2 comes from averaging over the full period of the Josephson oscillation, and a factor of 1/4 comes from the 1:1 current division between the Josephson source and a matched load. Bias of the transmitter junction above the gap leads to injection of quasiparticles in the junction leads; in the presence of reduced quasiparticle density $x_{\rm qp}$, we have 
\begin{equation}
    I_0 (x_{\rm qp}) = I_0(0)(1-x_{\rm qp}),
\end{equation}
where $I_0(0)$ is the junction critical current in the absence of quasiparticle poisoning \cite{Lenander11}. For the bias regime studied here, we find that quasiparticle poisoning leads to a suppression of transmitter junction critical current of order 10\%, with a weak dependence on bias voltage out to a Josephson frequency of $\sim$500~GHz.

At higher bias voltage, the reduced quasiparticle density in the junction leads increases sharply and eventually saturates due to quasiparticle-induced suppression of the energy gap. For bias of the transmitter junction at Josephson frequency $>500$~GHz, we observe an upturn in measured quasiparticle poisoning rate that is not captured by our antenna model. We believe that the enhanced pair-breaking at the qubit is due to the incoherent emission of recombination photons in the leads of the transmitter junction. Detailed modeling of this physics is the subject of ongoing work.

\subsubsection{Absorption by Receiving Antenna}
To calculate the rate of absorption of pair-breaking photons by the qubit, we need to know the circulating power in the shared aluminum enclosure of the transmit/receive experiment for a given applied bias (see Fig.~\ref{fig:AntennaCartoon}c). To perform a naive estimate, we consider the case where absorption of Josephson radiation in the aluminum walls of the shared enclosure is the dominant source of photon loss. The circulating power can then be obtained from a simple power balance: the power radiated by the transmitter (Eq. \ref{eq:P_rad_sup} above) must equal the power absorbed by the aluminum. We assume that photons from the transmitter will reflect many times within the enclosure prior to absorption and that these reflections randomize the direction and polarization of the Josephson radiation, creating an isotropic distribution of radiation with random polarization. The fact that the measured quasiparticle poisoning rate does not depend on separation and relative orientation of the transmitter and receiver devices lends support to this assumption (see Fig. \ref{fig:S_XmonParity}). 

The power $P_{\rm enc}$ absorbed by the aluminum enclosure is given by
\begin{equation}
    P_{\rm enc} = \pi \eta SA_{\rm enc}, 
\end{equation}
where $\eta=1-\left|\frac{Z_{\rm Al}-Z_{\rm fs}}{Z_{\rm Al}+Z_{\rm fs}}\right|^2$ is the efficiency with which photons are absorbed by the aluminum enclosure, $S$ is the incident power per unit solid angle per unit area, and $A_{\rm enc}$ is the inner area of the enclosure (the factor $\pi$ comes from integrating the cosine of the polar angle over the half-space). Here, $Z_{\rm fs}=377 ~\Omega$ is the impedance of free space and $Z_{\rm Al}$ is the surface impedance of aluminum:
\begin{equation}
    Z_{\rm Al} = (1+j)\sqrt{\frac{\omega \mu_0}{2\sigma_{\rm Al}}},
\end{equation}
where $\mu_0$ is the permeability of free space and where for $\sigma _{\rm Al}$ we take the conductance of aluminum 6061 at 4~K \cite{Clark1970}; see Table~\ref{tab:S_cal}.

For isotropic coherent radiation with wavelength $\lambda$, 
the power absorbed by the receiving antenna is
\begin{equation}
    P_{\rm rec} = \frac{1}{2}S \lambda^2 e_{\rm c, rec},
\end{equation}
where the factor $1/2$ accounts for random polarization of the radiation field. The rate of absorption of photons from the Josephson radiation field is then $\Gamma_{\rm J}(f) = P_{\rm rec}/hf$.

\subsubsection{Summary and Comparison with Measurement}
Collecting the various pieces, we arrive at the following expression for the expected rate of photon absorption at the receiver antenna, within our simple model:
\begin{equation}
    \Gamma_{\rm J, ~theo}(f) = \frac{e_{\rm c, tr}\,e_{\rm c, rec}}{16\eta} \frac{\lambda^2}{A_{\rm enc}}\frac{I_{0}^2R_{\rm n}}{hf}.
\end{equation}

In Table \ref{tab:S_cal}, we provide numerical values for the various quantities involved in this expression (the calculated coupling efficiencies $e_{\rm c,tr}$ and $e_{\rm c,rec}$ are presented in Fig.~\ref{fig:Xmon} of the main text). 

The measured rate parity $\Gamma_{\rm p}$ of parity switches can be broken into two parts:
\begin{equation}
    \Gamma_{\rm p}(f)=\Gamma_0+\Gamma_{\rm J}(f).
\end{equation}
Here, the baseline parity rate $\Gamma_0$ accounts for quasiparticle poisoning induced by broadband radiation from higher temperature stages of the cryostat, while $\Gamma_{\rm J}(f)$ is the contribution from Josephson radiation emitted by the transmitter junction.

In Table. \ref{tab:PATCompare} we compare the expected rate  $\Gamma_{\rm J, theo}$ of parity switches induced by Josephson oscillations to the measured rate $\Gamma_{\rm J, meas}$ at the frequencies where the spectral response of the qubit peaks. We find that the measured rates at 190~GHz (270~GHz) are a factor 10 (20) smaller than those expected from our naive calculation. This discrepancy could be due to additional photon losses in the materials of the enclosure, including Rogers PC board launchers, or to trapping of photons in the shallow cavity beneath the transmitter chip due to near-continuous coverage of the device layer by its Nb groundplane. 
%preferential emission of the transmitter in the direction opposite the receiver antenna, where photons could become trapped by by the Nb groundplane in the shallow cavity underneath the transmitter chip with 

\begin{table}[t]
\centering
\caption{\textbf{Parameters used for calculation of the expected rate $\Gamma_{\rm J, theo}$ of parity switches in the experiment of Fig. \ref{fig:Xmon}d.} The critical current without quasiparticle suppression is calculated from $R_{\rm n}$ using the Ambegaokar-Baratoff relation \cite{Ambegaokar82}; the suppression of $I_0$ due to local quasiparticle injection is extracted from a simple rate equation \cite{Wang14} using the reduced quasiparticle recombination rate (438 ns)$^{-1}$ \cite{Kaplan}.}
\begin{tabular}{ |c|c|c| } 
\hline
Parameter & value & source \\
\hline
Al conductivity $\sigma_{\rm Al}$ & $7.2\times 10^{7} ~\rm S/m$ &\cite{Clark1970}\\ 
\hline
Enclosure inner area  $A_{\rm enc}$ & $3.2\times10^{-3} ~\rm m^2$ &  design \\ 
\hline
Transmitter junction $R_{\rm n}$ & 33.0 k$\Omega$ &  RT 4-wire probe \\ 
\hline
Transmitter junction $I_{0}$ & 8.3 nA &  see caption \\ 
\hline
\end{tabular}
\label{tab:S_cal}
\end{table}

\begin{table}[t]
\centering
 \caption{\textbf{Transmit/receive experiment: comparison between naive model and measurement.}}

\begin{tabular}{ |c|c|c|c| } 
 \hline
  $f_{\rm J}$ & $\Gamma_{\rm J, ~\rm theo}$ & $\Gamma_{\rm J, ~\rm meas}$ & $\Gamma_{\rm J, ~\rm meas}/\Gamma_{\rm J, ~\rm theo}$ \\ 
 \hline
  $ 190 ~\rm GHz$ & $1020 ~\rm s^{-1}$  & $103 ~\rm s^{-1}$ & 0.10 \\ 
 \hline
   $ 270 ~\rm GHz$ & $ 11400 ~\rm s^{-1}$  & $528 ~\rm s^{-1}$ & 0.05 \\ 
 \hline
\end{tabular}
\label{tab:PATCompare}
\end{table}

\subsection{Extended Xmon Dataset}\label{sec:XmonEx}

\begin{figure}[t!]
\includegraphics[width=\columnwidth]{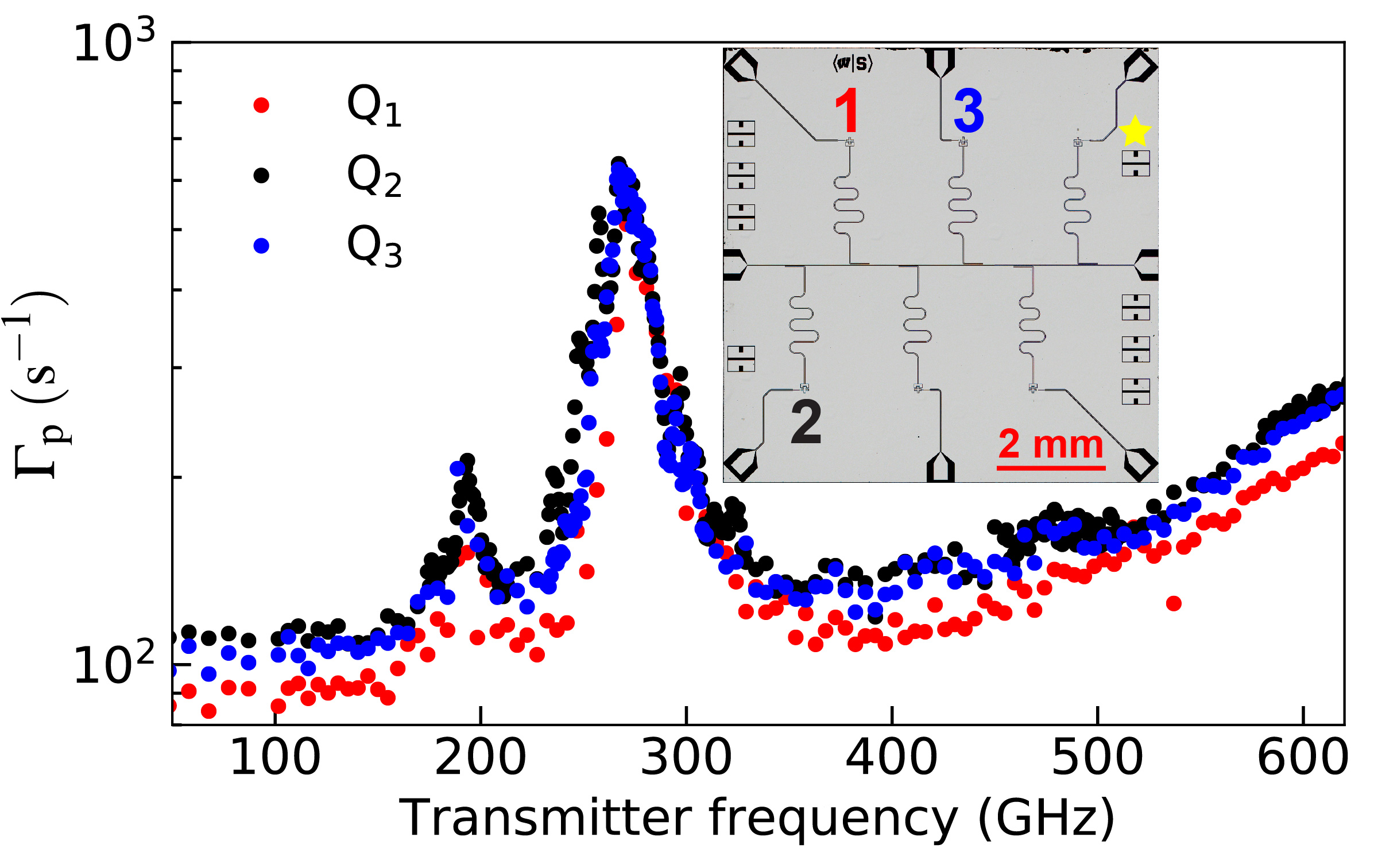}
\caption{\textbf{Spectral response of three Xmon qubits.}
Three qubits of similar design at different positions are measured in the same cooldown. They share the same transmitter (Fig. \ref{fig:Xmon}a). Extracted parity rates show similar response to the transmitter. Inset shows the positions of the three qubits on the receicer chip. The yellow star shows the relative location of the transmitting antenna on the other chip. The two chips are positioned face to face with 9.6 mm chip to chip spacing shown as in Fig. \ref{fig:AntennaCartoon}c.}
\label{fig:S_XmonParity}
\end{figure}

We measured the rate of parity switches induced by Josephson radiation from a single transmitter junction for three different Xmon devices on the same receiver chip. The receiver chip geometry is shown in the inset of Fig.~\ref{fig:S_XmonParity}a. The receiver and transmitter chips are mounted  face to face with a separation of 9.6 mm; in this inset, the position of the transmitter junction (as seen looking through the chips) is indicated by a yellow star. The three Xmons share similar island geometry and junction parameters (see Table \ref{tab:qparams} for details; one of the qubits has a slightly smaller island corresponding to a slightly higher fundamental antenna resonance frequency). In Fig. \ref{fig:S_XmonParity}a, we plot the parity rates measured for these three devices as a function of Josephson frequency of the transmitter. We observe similar resonant structure in all devices, despite different separation and relative orientation with respect to the transmitter. This observation lends support to the assumption that multiple reflections in the shared aluminum enclosure will randomize the direction and polarization of the Josephson radiation. 

\begin{figure}[t]
\includegraphics[width=\columnwidth]{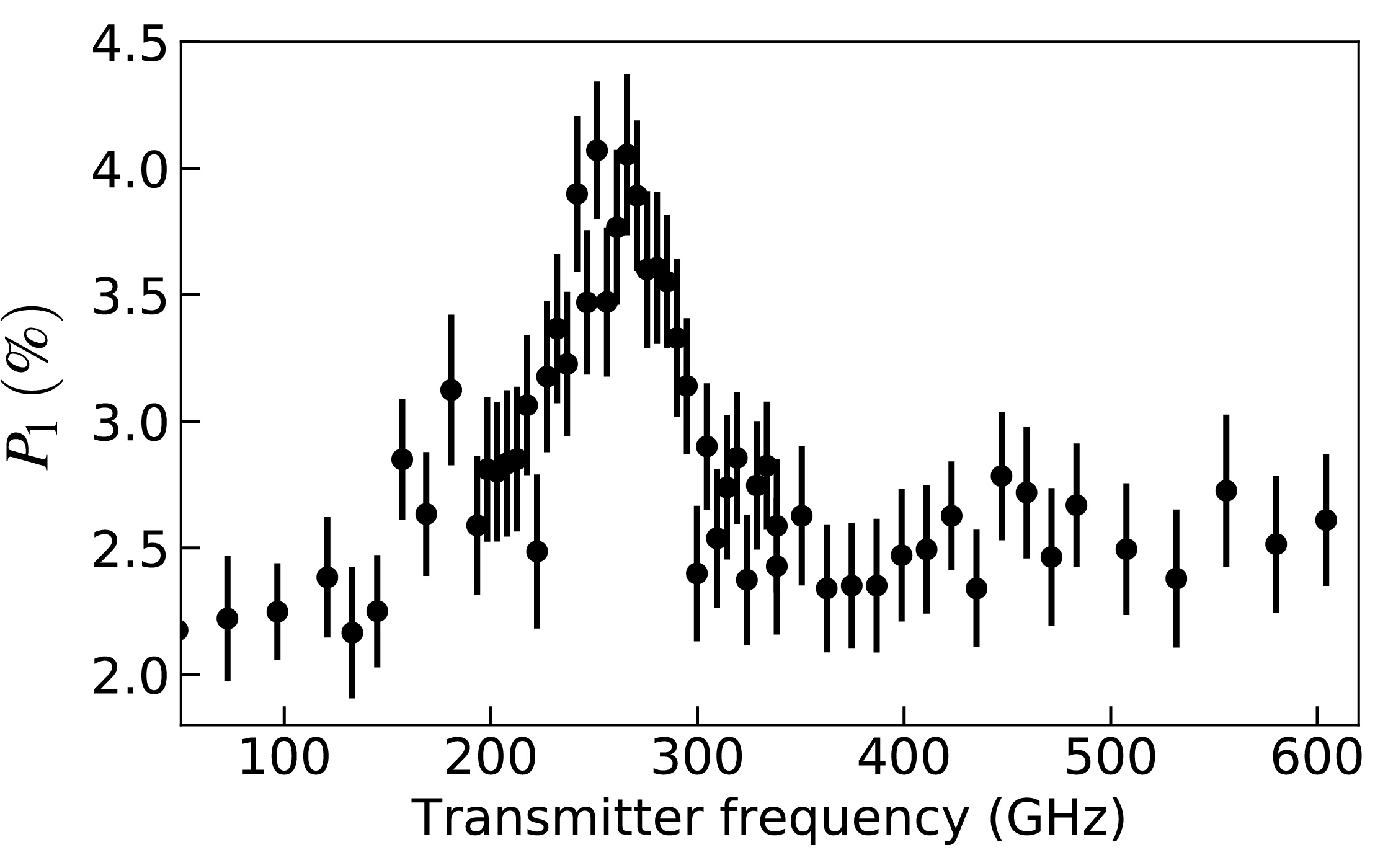}
\caption{\textbf{Dependence of $\ket{1}$ occupation on transmitter frequency.}
Occupation of the qubit $\ket{1}$ state is enhanced due to resonant absorption of pair-breaking photons. The peak at 270~GHz coincides with the peak in photon transfer efficiency from the transmitter junction to the Xmon qubit.
}
\label{fig:S_XmonP1}
\end{figure}

In a separate experiment, we measure the equilibrium $\ket{1}$ state occupation of Xmon $Q_2$ as a function of transmitter frequency. Here we use a more strongly coupled transmitter with the same geometry as that shown in Fig.~\ref{fig:Xmon}a, but with different junction parameters corresponding to a factor 7.9 higher critical current and a factor 1.6 greater coupling efficiency at 270~GHz. In Fig. \ref{fig:S_XmonP1}, we plot $P_{1}$ versus transmitter frequency. While this measurement is far less sensitive than the Ramsey-based parity experiment, we see that the excess $\ket{1}$ state occupation displays similar resonant structure due to qubit transitions induced by absorption of pair-breaking photons. 

\subsection{Correlation between Qubit Transitions and Parity Switches}\label{sec:PUp}
\begin{figure}[t!]
\includegraphics[width=\columnwidth]{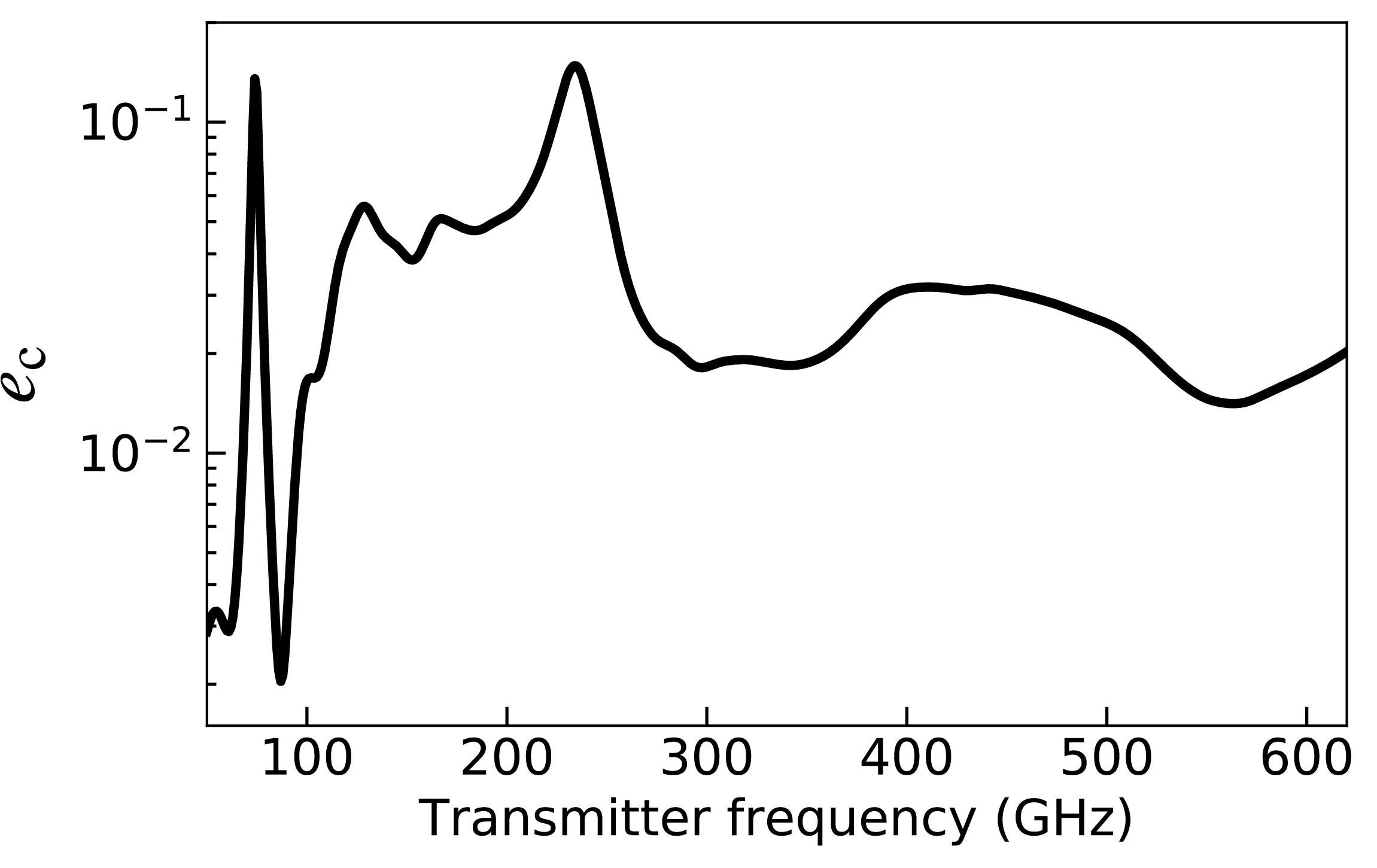}
\caption{\textbf{Coupling efficiency of transmitter used in the experiments of Fig.~\ref{fig:Up} in the main text.}}
\label{fig:S_transmit}
\end{figure}

\begin{figure}[t!]
\includegraphics[width=\columnwidth]{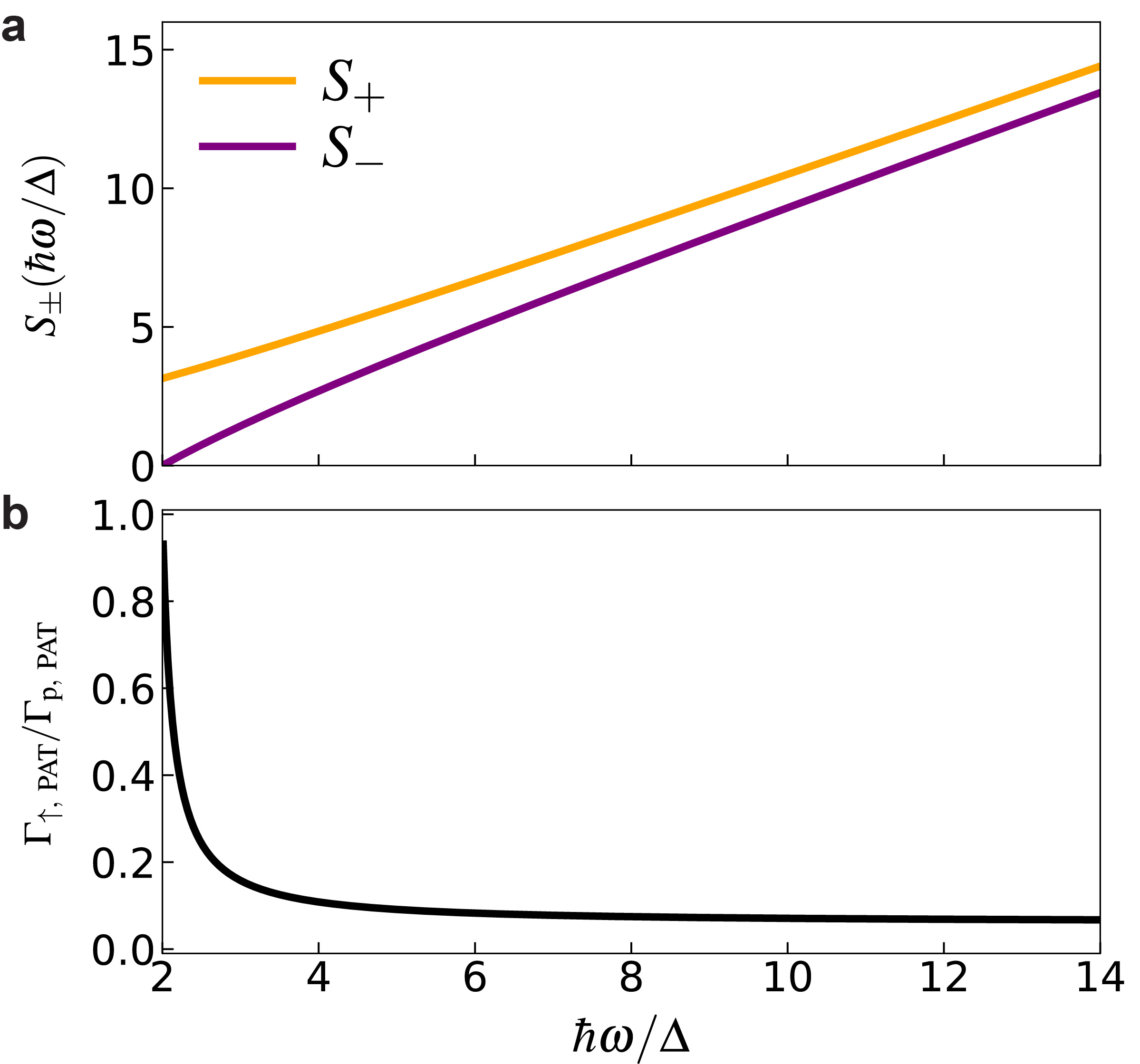}
\caption{
\textbf{a}, Structure factors for photon-assisted quasiparticle poisoning events. 
\textbf{b}, Ratio $\Gamma_{\uparrow, ~\rm PAT}/\Gamma_{\rm p, ~\rm PAT}$ of spurious qubit upward transitions to quasiparticle poisoning events induced by photon absorption as a function of photon energy. Here, we take $E_{\rm J}/E_{\rm c} = 28$.
}
\label{fig:S_pm}
\end{figure}

The absorption of a pair-breaking photon at the qubit junction can induce qubit state transitions; a detailed analysis of this physics has been performed by Houzet \textit{et al.} \cite{Houzet19}. We have separately characterized the rate of parity switches $\Gamma_{\rm p, ~\rm PAT}$ and upward qubit transitions $\Gamma_{\uparrow, ~\rm PAT}$ induced by coherent mm-wave irradiation from the transmitter; see Fig.~\ref{fig:Up} in the main text. The coupling efficiency of the transmitter junction used in these experiments is shown in Fig.~\ref{fig:S_transmit}. From \cite{Houzet19}, the rate of transitions out of the qubit $\ket{0}$ state and the rate of parity switches induced by photon absorption are related as follows:
\begin{equation}\label{eq:PAT1}
    \Gamma_{\uparrow, ~\rm PAT} = \left(1+\sqrt{\frac{8E_{\rm J}}{E_{\rm c}} \frac{S_{-}}{S_{+}}}\right)^{-1}\Gamma_{\rm p, ~\rm PAT},
\end{equation}
where the structure factors are defined as 
\begin{equation}
S_{\pm} (\hbar\omega/\Delta)=\int _{1} ^{\hbar\omega/\Delta-1}dx\frac{x(\hbar\omega/\Delta-x)\pm 1}{\sqrt{x^2-1}\sqrt{(\hbar\omega/\Delta-x)^2-1}};
\end{equation}
see Fig.~\ref{fig:S_pm}a.

In Fig.~\ref{fig:S_pm}b, we plot the ratio $\Gamma_{\uparrow, ~\rm PAT}/\Gamma_{\rm p, ~\rm PAT}$ obtained from Eq.~\ref{eq:PAT1} versus energy of the pair-breaking photon. Here we take $E_{\rm J}/E_{\rm c} = 28$, as relevant for circmon device $Q_2$ studied in Fig.~\ref{fig:Up} of the main text. For high photon energy, we see that this ratio saturates to 6\%. However, the measured ratio of upward transitions to parity switches is 20\%. The enhanced rate of spurious transitions out of the $\ket{0}$ state can be explained in terms of additional qubit transitions induced by diffusion of quasiparticles across the qubit junction following the primary poisoning event.

\end{document}